\documentstyle[12pt,aasms4]{article}
\begin{document}

% Next 5 lines define \simless and \simgreat: "less than or approximately
% equal to" and "greater than or approximately equal to".
\newbox\grsign \setbox\grsign=\hbox{$>$} \newdimen\grdimen \grdimen=\ht\grsign
\newbox\simlessbox \newbox\simgreatbox
\setbox\simgreatbox=\hbox{\raise.5ex\hbox{$>$}\llap
     {\lower.5ex\hbox{$\sim$}}}\ht1=\grdimen\dp1=0pt
\setbox\simlessbox=\hbox{\raise.5ex\hbox{$<$}\llap
     {\lower.5ex\hbox{$\sim$}}}\ht2=\grdimen\dp2=0pt
\def\simgreat{\mathrel{\copy\simgreatbox}}
\def\simless{\mathrel{\copy\simlessbox}}

\title{THE TEMPERATURE SCALE OF METAL-RICH M GIANTS BASED ON TiO BANDS: 
POPULATION SYNTHESIS IN THE NEAR INFRARED
 \altaffilmark{1}}
\altaffiltext{1}{Observations obtained at the
Laborat\'orio Nacional
de Astrof\'\i sica (LNA), Brazil, and European Southern Observatory
(ESO), Chile}

\author{\bf R. P. Schiavon\altaffilmark{2,3}, 
B. Barbuy\altaffilmark{2}}

\altaffiltext{2}{Universidade de S\~ao Paulo, IAG, Departamento de Astronomia,
C.P. 3386, S\~ao Paulo 01060-970, Brazil,
Email: ripisc@on.br, barbuy@orion.iagusp.usp.br}

\altaffiltext{3}{Present Address: CNPq/Observat\'orio Nacional, 
Departamento de Astronomia, Rua
General Jos\'e Cristino, 77, Rio de Janeiro, 20921-400, Brazil.}

\slugcomment{Submitted to The Astrophysical Journal}
\slugcomment{Send proofs to:  R. P. Schiavon}

\begin{abstract}
We have computed a grid of high resolution synthetic spectra
for cool stars (2500 $<$ T$_{eff}$ $<$ 6000 K) in the
wavelength range $\lambda\lambda$ 6000 -- 10200 {\rm \AA}, by 
employing an updated line list of atomic and molecular lines,
together with state-of-the-art model atmospheres.

As a by-product, by fitting TiO bandheads in spectra of well-known M
giants, we have derived the electronic oscillator strengths of the TiO
$\gamma$', $\delta$, $\epsilon$ and $\phi$ systems. The derived oscillator
strenghts for the $\gamma'$, $\epsilon$ and $\phi$ systems differ from
the laboratory and ab initio values found in the literature, 
but are consistent with the model
atmospheres and line lists employed, resulting in a good match to the
observed spectra of M giants of known parameters.

The behavior of TiO bands as a function of the stellar
parameters T$_{eff}$, log g and [Fe/H] is presented and the use of TiO
spectral indices in stellar population studies is discussed. 
\end{abstract}

\keywords{Atomic and molecular data, Stars: atmospheres, M giants,
globular clusters} 

\section{INTRODUCTION}
 
A series of TiO spectral features are present in the integrated spectra
of normal galaxies. Measurements
 of TiO$_1$ and TiO$_2$ indices as
defined in Burstein et al. (1984) and Worthey et al. (1994),
and several indices given in Bica \& Alloin (1986) are available
in the literature. However,  the  use of TiO bands as
stellar population indicators is not widespread.

One clear characteristics of TiO bands is the dependence of
their intensity on effective temperature.
The effective temperature (T$_{\rm eff}$) scale of M giants has been the
subject of many studies in the past, both from the theoretical and
observational standpoints -- see the recent work by Bessell et al. (1998,
hereafter BCP98) and Lejeune et al. (1998)
The main source of uncertainty in the temperature scale of M giants ir
related to uncertainties in the stellar 
mass, surface gravity and metallicity (Alvarez \& Plez 1998).

In this work, we use a revised atomic and molecular line list 
and  state-of-the-art
model atmospheres to synthesize the spectra of M giants in the 
near-infrared (NIR). A comparison of synthetic spectra with observations of cool 
red  giants permits a calibration of T$_{\rm eff}$s for cool giants of 
the bulge metal-rich globular cluster NGC
6553. 

A grid of synthetic spectra in the wavelength range 
$\lambda\lambda$ 6000 -- 10200 {\rm \AA} is computed and equivalent widths
of TiO bands as a function of stellar parameters (T$_{\rm eff}$, log g, [Fe/H]) 
are
measured.

In Sec. 2 the observations are reported. In Sec. 3
the calculation of  synthetic spectra is described. In
Sec. 4 we derive  effective temperatures for M giants of the
globular cluster NGC 6553, 
based on the intensity of NIR TiO bands, and we derive
electronic oscillator strengths for the  
TiO $\gamma$', $\delta$, $\epsilon$ and $\phi$ systems.
In Section 5 the computation of a grid of synthetic spectra in the
NIR  and measurements of equivalent widths of TiO features
are presented; using the grid of synthetic spectra, the composite 
spectrum for a single-age stellar population is built and
compared to the observed integrated spectrum of the globular
cluster NGC 6553. In Section 6
conclusions are drawn.

\section{ OBSERVATIONS}

We have observed 6 stars among the coolest giants of the  bulge metal-rich
globular cluster NGC 6553 and two well-known field M giants HR 625 and HR
3816.  The selection and identification of the stars in NGC 6553 were based on
the images 
and Colour-Magnitude Diagrams (CMDs) 
given in Ortolani et al. (1990) and Guarnieri et al.
(1998). The list of stars is shown in Table 1.

The spectra of individual stars from the RGB tip of NGC 6553 were
collected at the Cassegrain focus of the 1.6m telescope at the {\it
Laborat\'orio Nacional de Astrof\'\i sica} (LNA), Pico dos Dias,
Brazil, in 1997 June, using a EEV CCD of 770$\times$1152 pixel (LNA CCD
\# 48) of pixel size 22.5$\mu$x22.5$\mu$m, and a grating of 830l/mm,
with blaze angle at 7490 {\rm \AA}. This combination gives a spectral
resolution of $\Delta\lambda$ $\sim$ 3 {\rm \AA}, centered at
$\lambda$8000 {\rm \AA} with a spectral coverage of $\Delta\lambda$
$\sim$ 1350 {\rm \AA}. In order to enlarge the spectral interval to
include bands from all the relevant electronic transitions of TiO, we
collected two spectra for each star, centered at $\lambda$7570 and
$\lambda$8235 {\rm \AA} with an overlap in the interval
$\lambda\lambda$7570--8250 {\rm \AA}.  At the end of the reduction
process, the spectra were combined and the final spectral interval is
$\lambda\lambda$ 6865--8905 {\rm \AA}.  The flux calibration shows a
systematic turnover around $\lambda$ 8700 {\rm \AA}, which affects the
spectrum shape at the $<$ 5\% level. The effect of this photometric
inaccuracy on relative band intensities is negligible.

An integrated spectrum of NGC 6553 and the M giant HR 625 were observed
at LNA in 1996 July and in 1996 August respectively, at the Cassegrain
focus of the 1.6m telescope, using a grating of 300l/mm, with blaze
angle at $\lambda$10000 {\rm \AA}.  Employing a 1024$\times$1024 pixels
thin back-illuminated SITe CCD (LNA CCD \# 101), of pixel size
24$\mu$mx24$\mu$m, a spectral coverage $\lambda\lambda$ 6020 -- 10380
{\rm \AA}, with an effective resolution of 8{\rm \AA}, was obtained.

The M giant HR 3816 was observed in 1996 February at the 1.5m telescope 
of the {\it European
Southern Observatory} - ESO
using the Ford Aerospace ESO CCD \# 24, of 2048x2048 pixels, with
pixel size 15$\mu$mx15$\mu$m,
together with a grating of 1200 l/mm (ESO grating \# 11), giving
a dispersion of 66 {\rm \AA}/mm or a spectral resolution of
$\Delta\lambda$ $\sim$ 2 {\rm \AA}.

Reductions were carried out using IRAF, following the usual steps: bias
subtraction, flat-field and illumination correction, spectrum
extraction and wavelength calibration. The spectra were flux calibrated
through observations of three spectrophotometric standards per night,
taken from Hamuy et al. (1994). Telluric lines have been removed
through division by spectra of high $v\sin i$  B stars.

\section{ SPECTRUM SYNTHESIS}

A grid of synthetic spectra in the wavelength range $\lambda\lambda$
6000 -- 10200 {\rm \AA} is computed, using a revised version (Barbuy,
B., Perrin, M.-N., in preparation) of the code described in Barbuy
(1982), which incorporates the computation of molecular lines to the
code for atomic lines by Spite (1967), assuming local thermodynamic
equilibrium.  The elemental abundances adopted are from Grevesse et
al.  (1996).  The atomic lines are identified in the solar spectrum and
oscillator strengths are obtained through a fit of synthetic spectra to
the solar spectrum (Delbouille et al.  1973).  The molecular systems
taken into account are the CN  (A$^2\Pi$--X$^2\Sigma$) red system,
C$_2$ (A$^3\Pi_g$--X$^3\Pi_u$)  Swan system, TiO
$\gamma$(A$^3\Phi$--X$^3\Delta$), $\gamma'$(B$^3\Pi$--X$^3\Delta$),
$\delta$(b$^1\Pi$--a$^1\Delta$), $\epsilon$(E$^3\Pi$--X$^3\Delta$) and
$\phi$(b$^1\Pi$--d$^1\Sigma$) systems, and the FeH
(A$^4\Delta$--X$^4\Delta$) electronic system, which contains the
Wing-Ford band.

The detailed study of the C$_2$ Swan, CN red and TiO $\gamma$ systems
and respective molecular constants employed can be found in Barbuy
(1985), Erdelyi-Mendes \& Barbuy (1989), Barbuy et al.  (1991), and
Milone \& Barbuy (1994).  The list of FeH lines is described in
Schiavon et al. (1997).  The line lists and corresponding molecular
constants for the TiO $\gamma'$, $\delta$, $\epsilon$ and $\phi$
systems, adopted from Jorgensen (1994), are discussed in Section 4.

The photospheric models employed are from Kurucz (1992), Plez et al.
(1992) and further unpublished models by Plez (1997), and Allard \&
Hauschildt (1995). The values of effective temperatures, surface
gravities and metallicities adopted in the grid of models are reported
in Table 4 (Sec. 5). Microturbulent velocities adopted were ${\rm
v_{\rm t}}=1.0km/s$ and 2.5km/s for $\log g \geq$ 3.0 and $\log g \leq$
2.5 respectively. The Plez et al. (1992) grid, which was employed in
the computation of spectra in the range 2500 $\leq$T$_{\rm eff}\leq$
3400 K is only available for [Fe/H]=0.0, so that for [Fe/H] = --0.3 and
--0.6 we have used the models for [Fe/H]=0.0, and the nonsolar
metallicities were taken into account only in the input elemental
abundances.

The calculations were carried out in steps of 0.02{\rm \AA} and the
resulting spectra were rebinned to a step of 1{\rm \AA} and convolved
with a gaussian function simulating an instrumental profile of
FWHM=2{\rm \AA}.  We note that for the M giants studied here, it is
important to take into account, in the calculation of continuum
opacity,  the H$_2$ formation, which reduces the opacity due to HI and
HII species.

The photospheric C and N values  are modified in giants due to convective mixing
so that we adopted [C/Fe] = --0.2 and [N/Fe] = +0.4 for giants of log g
$\leq$ 2.0,  and solar ratios for higher gravity stars. The change
 in C and N affects the TiO and CN bands: CN bands become stronger
and, because of C deficiency, TiO bands also become stronger, since
there is less CO association.

\section{ TiO BANDS AND THE T$_{eff}$ SCALE OF M GIANTS}

%The T$_{\rm eff}$ scale of M giants is a long standing problem. 

TiO bands are the main opacity source in the atmospheres of stars
cooler than 4000K in the spectral region under study, and M giants are
the dominant contributors to the integrated light of metal-rich and old
composite systems such as bulge globular clusters and normal galaxies
in the NIR.

In Figure 1 are shown the contributions of the most important
electronic systems of TiO in the studied spectral region,
 for a star of T$_{\rm eff}$=3200K, $\log g$=0.0 and [Fe/H]=--0.3.
 For $\lambda \simless$ 8000 {\rm \AA},
the opacity is dominated by the $\gamma$ and $\gamma'$ systems, whereas
for $\lambda \simgreat$ 8000 {\rm \AA},
 the $\delta$ and $\epsilon$ systems are dominant, the opacity
due to the $\phi$ system being important only for $\lambda\simgreat$
10000 {\rm \AA}.

The synthesis of the $\gamma$ system was already presented in  Milone
\& Barbuy (1994): the laboratory line list by J.G. Phillips (see
Phillips 1973) is used, and H\"onl-London and Franck-Condon factors
were computed following the equations given by Kovacs (1979) and Bell
et al. (1979) respectively; the electronic oscillator strength $f_{el}$
adopted here is given in Table 3 and discussed in Sec. 4.2.  We have
chosen to use the laboratory line list because it reproduces better the
shape of the (0,0) $\gamma$ system  bandhead (see discussion in Sec.
4.2.1).

The line lists used for the computation of the $\gamma'$, $\delta$, $\epsilon$
and $\phi$ systems, including the normalized product of the Franck-Condon and
H\"onl-London factors, were the ones computed by Jorgensen (1994),  which were
kindly made available to us by R. Bell. For these systems, while the 
H\"onl-London and Franck-Condon factors were adopted directly from Jorgensen
(1994), we derived eletronic oscillator strengths  $f_{el}$ values by fitting
the main vibrational bands in the spectra of M giants. 

The fit to the intensity of TiO bandheads was applied to M stars of 
known stellar parameters (T$_{\rm eff}$, log g, [Fe/H], v$_{\rm t}$): 
two well-known field M giants and M giants
 along the cool red giant branch of the metal-rich globular 
cluster NGC 6553.
We searched for VO bands in our spectra by following
identifications by Barnbaum et al. (1996), and found no
evidence for the presence of these bands, even for the coolest
stars of our sample.
The determination of the oscillator strengths proceeded in two steps, described
in detail in the subsections below:
(a) fine-tuning of the effective temperature of cool giants of NGC 6553,
based on their TiO $\gamma$ system bandheads;
(b) derivation of electronic oscillator strengths of
the $\gamma$', $\epsilon$, $\delta$ and $\phi$ systems,
by fitting their main bandheads, adopting the temperatures
established in (a). 

\subsection {Stellar parameters for M giants of NGC 6553}

In a detailed analysis of individual stars of NGC 6553,
 Barbuy et al. (1997) obtained a metallicity [Fe/H] = --0.35
and a titanium-to-iron abundance ratio of [Ti/Fe] $\approx$ +0.3.
In a further analysis of these data, Barbuy et al. (1998) have
confirmed that, in spite of a temperature uncertainty of 200 K,
[Ti/H] $\approx$ 0 is found in all cases. As also discussed in
Bruzual et al. (1997), a metallicity of [Fe/H] = --0.3 to --0.4
combined to enhanced [$\alpha$-elements/Fe] = +0.3 to +0.4
result in Z/Z$_{\odot}$ $\approx$ 1 or [M/H] $\approx$ 0.
Furthermore, given that in our calculations the TiO bands
are the dominant feature, and that [Ti/H] $\approx$ 0 is found for
NGC 6553, it is clear that an overall metallicity [M/H] = 0.0
is to be used. In conclusion, we adopt [Fe/H] = 0.0 for 
NGC 6553, where by [Fe/H] we mean in fact [M/H]. 

For our sample of M giants, 
temperatures are first derived from the (V--K) colours given in
Guarnieri et al. (1998). In Table 2 are given the observed (V--K), 
for which
reddening corrections were applied using E(B--V) = 0.7 and 
E(V--K)/E(B--V) = 2.744 (Rieke \& Lebofsky 1985),
and temperatures derived through the calibration
 of Ridgway et al. (1980), LCB98 and BCP98.

The $\gamma$ system bandheads were used to establish our T$_{\rm eff}$ scale
for the cool stars, given that among the TiO systems considered in this
work,  it presents the best settled molecular constants available. 

\subsubsection{Fits to reference stars}

The fit to the field M giant HR 625, shown in Fig. 2, resulted in
stellar parameters (T$_{\rm eff}$, log g, [Fe/H]) = (3600, 1.0, 0.0).
This T$_{\rm eff}$ is $\sim$150K lower than the one estimated by
Schiavon et al. (1997), on the basis of the spectral type M2 (Hoffleit
\& Warren, 1991) and the calibration by Fluks et al. (1994).  In fact,
the fit of the FeH bands in this star, adopting T$_{\rm eff}$ = 3750 K
was not very satisfactory in Schiavon et al. (1997).

The fit to the vibrational sequence $\Delta$v = 0 (v',v'') = [(0,0), (1,1),
(2,2), (3,3), etc.] 'composite  bandhead' of the $\gamma$ system at $\lambda$
7150 {\rm \AA} to the spectra of stars in NGC 6553 (Table 2), was applied in a
fine-tuning of their effective temperatures,  which are given in the column 8
of Table 2. For
each T$_{\rm eff}$ we adopted a surface gravity based on the T$_{\rm
eff}\times\log g$ relation given by LCB98, which is based on the evolutionary
tracks of Schaller et al. (1992) for 1 M$_\odot$ (column 9 of Table 2).

The T$_{\rm eff}$ estimations given in Table 2 indicate that (i)
temperatures from the (V--K) empirical fits of BCP98 are lower than
those derived from the theoretical values of their Table 5; (ii)  our
TiO-based effective temperature estimations show a good agreement with
the resulting temperatures from  (V--K) calibrations by Ridgway et al.
(1980), LCB98 and BCP98 (empirical). The T(TiO) values are adopted
here. Taking into consideration the uncertainties in metallicity and
$\log g$, we estimate an error of $\Delta$T$_{\rm eff}$=150K for the
stars hotter than T$_{\rm eff}\sim$3200K, for which the gravity effect
on band intensities is minimmum (see discussion on Sect. 5). For cooler
stars, gravity effects are more important, so that uncertainties can
reach $\Delta$T$_{\rm eff}$=250K. Let us note that for T$_{\rm eff}$
$\simless$ 3200 K, the TiO $\gamma$ $\Delta$v=0 bandhead becomes
saturated, so that for the star 291 (V12) the derived temperature is
less precise - see also discussion on saturation of TiO bands in Sec.
5.

In Fig. 3 the fit of a  synthetic spectrum
of the $\Delta$v=0 vibrational sequence of the $\gamma$ 
system to the star 
288 of NGC 6553 (T$_{\rm eff}$ = 3400 K) is shown.

\subsection  { Electronic oscillator strengths for the TiO
$\gamma'$, $\epsilon$, $\delta$ and $\phi$ systems}
\par 

The final $f_{el}$ values obtained are given in Table 3, together with
other determinations collected from the literature: the values
 by Davis  et al. (1986) were obtained from correlations between
equivalent widths of rotational lines and lifetime estimates for the
$\alpha$ and $\beta$ systems; Brett (1990) obtained semi-empirical
values through the fit of observed spectra of M giants;
 Jorgensen (1994) reports $f_{el}$ values used in his calculations;
Hedgecock et al. (1995) derived $f_{el}$ from lifetime measurements;
Langhoff (1997) ones were obtained from {\it ab initio} calculations;
Alvarez \& Plez (1998) present updated f$_{el}$ values from recent
laboratory measurements or calculations, which they also checked
through comparisons between observed and calculated spectra of M
giants. 

The $f_{el}$ values used in our calculations were derived in this work
as follows:

(a) Adopting the T$_{\rm eff}$s as described in Sec. 4.1,  we derive
$f_{el}$ values for the $\epsilon$ and $\delta$ systems, by fitting the
observed intensity of their $\Delta$v=0 vibrational bands, as shown in
Fig. 4. In the case of the $\epsilon$ system, we had to shift the
wavelengths of all lines by $\Delta\lambda = $ +76.0 {\rm \AA}
 in order to correctly reproduce the position of the (0,0) bandhead at
$\approx$ $\lambda$8400 {\rm \AA}.  This $\lambda_0$ shift is
compatible with an uncertainty of the order of 100  cm$^{-1}$ in the
electronic term T$_{\rm e}$ of the upper level (E$^3$$\Pi$) of the
transition (see Jorgensen 1994; Huber \& Herzberg 1979; Simard \&
Hackett 1991). We find T$_{\rm e}$ = 11829.3 for the E$^3$$\Pi$ level,
by shifting Jorgensen's value.  The value obtained for
this system (f$_{el}$ = 0.07) is in disagreement with all values in the
literature by a factor $>$ 10. Recent laboratory determinations by
Lundevall (1998) give a lifetime of 4.0$\pm$0.2 $\mu$s for
the E$^3$$\Pi$ level, wherefrom a f$_{el}$ even lower than Langhoff's
would be derived. Therefore we preferred to omit our result for the
$\epsilon$ system in Table 3.  For our purposes of computing a library
of synthetic spectra, it is only important that the combination of line
list, Franck-Condon factors, H\"onl-London factors and f$_{el}$ match
the observed spectra.  We suggest 
 that the line list of Jorgensen (1994) for the $\epsilon$
system is to be revised.

(b) The intensity of the vibrational sequences given by $\Delta$v = 0
[(v',v")=(0,0), (1,1), (2,2), (3,3), etc.] and $\Delta$v = --1
[(v',v")=(0,1), (1,2), (2,3), etc.]
 of the $\gamma$' system, at $\approx$ $\lambda$6200 and $\lambda$6700
{\rm \AA} respectively, were used to obtain its electronic oscillator
strength $f_{el}$, by fitting the spectrum of HR 625.

(c) The $f_{el}$ of the $\phi$ system was adopted from the ratio of the
theoretical $f_{el}$s derived by Langhoff (1997) for the $\delta$ and
$\phi$ systems, giving $f_{el}^\phi/f_{el}^\delta \sim$ 0.2 .

We adopted the procedure of fitting the bands from the different
systems to the spectrum of the star NGC6553-288, for which the T$_{\rm
eff}$ determined from the intensity of the bands from the $\gamma$
system is in best agreement with the T$_{\rm eff}$s derived from the
various colours. The f$_{el}$s so derived for the other systems
resulted in good fits to the spectra of all the remaining stars, of
Table 2, adopting T(TiO).

The values derived in this work are in most cases similar to the
average of the values estimated by other authors. The agreement between
different authors is best for the $\delta$ system.  For the $\gamma$
system, our adopted value of f$_{el}$ = 0.12 corresponds
essentially to a mean of the laboratory values f$_{el}$ = 0.165 and
0.092 by Davis et al. (1986) and Langhoff (1997).

\subsubsection {Errors in the electronic oscillator strenghts}

The main sources of uncertainty in the derivation of our astrophysical
f$_{el}$s are the uncertainties in T$_{\rm eff}$, $\log g$, Ti and
O abundances and the adopted microturbulent velocity (${\rm v_{\rm
t}}$). Another potential source of uncertainties is a possible
incompleteness in the laboratory line list adopted for the $\gamma$
system. The latter is an important issue given that the f$_{el}$s
of all the remaining systems are tied to the T$_{\rm eff}$s derived
from the intensities of the bands of the $\gamma$ system.

We have adopted a fixed  ${\rm v_{\rm t}}= 2.5km/s$ for giants. For
${\rm v_{\rm t}}=1.5km/s$ there would occur an increase in the
estimated f$_{el}$ of 15\%.

The uncertainty in the C and N abundances is estimated by computing the
$\gamma$ band with [C/Fe] = [N/Fe] = 0.0: the change relative to the
adopted values of [C/Fe] = --0.2 and [N/Fe] = 0.4 is of the order of
2\% in the estimated  f$_{el}$ value.

Adopting either a T$_{\rm eff}$ or a metallicity differing in 200K and
0.3dex respectively results in a shift of 20\% in the resulting
f$_{el}$s.  As the $\log g$ sensitivity of the TiO bands at T$_{\rm
eff}\sim$3400K is very low (see Sec.5), the influence of this parameter
upon the derived f$_{el}$s is very weak. An error of 0.5dex in $\log g$
results in a difference of only 4\% in the resulting f$_{el}$s.  Taking
into consideration all the above sources, the final uncertainty in our
f$_{el}$ values is around 30\%.

We also computed synthetic spectra using the theoretical line list from
Jorgensen (1994) for the $\gamma$ system, in order to check for a
possible incompleteness of the Phillips laboratory line list. As a
result, the f$_{el}$ that should be adopted with the Jorgensen line
list is only marginally lower (8\%) than the one that is compatible
with the Phillips line list. Taking into consideration the other
sources of uncertainty discussed above, we conclude that such an
incompleteness in the Phillips line list has only a mild effect in our
analysis. Moreover, we chose to adopt the Phillips line list because it
gives a better reproduction of the band shapes. The Jorgensen line list
gives too high opacities at high J and too low opacities near the
bandheads, so that it is very difficult to correctly reproduce the
whole observed band shapes when employing this line list.

The use of the f$_{el}$ value by Langhoff (1997) or 
Hedgecock et al. (1995), which are lower than ours by
respectively 30\% and 50\%, would imply too low temperatures
in comparison to those derived from the photometry (see below).
We consider that these values are incompatible with our data.

\section { A GRID OF SYNTHETIC SPECTRA: TiO vs. STELLAR PARAMETERS}

We have computed a grid of synthetic spectra for the stellar parameters
indicated in Table 4. A list of indices which measure TiO bands, as
detailed in Table 5, was measured for the whole grid of spectra, in
order to study their behavior as a function of stellar
parameters.

In Fig. 5 are shown the plots of equivalent widths of TiO bands
measured (cf. Table 5), for different $\log g$ and for a fixed metallicity
[Fe/H] = --0.3, as a function of T$_{\rm eff}$. 
The bandheads at $\lambda$ 6000, 6600, 7150, 7600 and 8400
{\rm \AA} show
similar behaviors, as expected. The dependence on gravity is also
shown.
In Fig. 6 the behavior of the indices, for a fixed $\log g$ = 0.0,
and different [Fe/H], as a function of T$_{\rm eff}$, shows the
strong dependence on metallicity.

The TiO$_2$ index as defined in Burstein et al. (1984)  (Figs.  5b and
6b) is much less sensitive to surface gravity than EW$_{6000}$.  In
order to explain this,   we plot in Fig. 7 the synthetic spectra for
Teff = 3400 K, log g = 0.5 and [Fe/H] = 0.0, where the TiO$_2$ index as
defined by Worthey et al. (1994) (the index is slightly modified
relative to Burstein et al. 1984), as well as our EW$_{6000}$ index are
indicated (see also Table 5). It can be seen that the TiO$_2$ index
measures the bottom of the TiO $\Delta$v=0 vibrational sequence of the
$\gamma$' system, where line saturation is stronger.  The index
is clearly  saturated  for T$_{\rm eff}$ $\simless$ 3500 K (for [Fe/H]
= 0.0), so that its sensitivity to both T$_{\rm eff}$ and $\log g$ is
very small in this T$_{\rm eff}$ range -- see Fig. 6b.

Figs. 5a-f and 6a-f show that the redder the TiO feature,
two effects occur: the threshold of detection of the bands
starts at progressively lower temperatures, and the bandheads
saturate for cooler temperatures. For example, EW$_{6000}$
and EW$_{8400}$ rise in strength for 
T$_{\rm eff}$ $\simgreat$ 4000 and 3400 K 
whereas saturation occurs at T$_{\rm eff}$ $\simless$
3000 and $\simless$ 2750 K  respectively.
It has to be noted that in all plots of Figs. 5 and 6 the equivalent
width corresponding to T$_{\rm eff}$=2500 K is underestimated, due
to the use of log g = --0.5, instead of log g = --1.0,
more appropriate for this temperature, since no models
are available for log g = --1.0 in the Plez grid.

One conclusion that can be drawn from these plots is that for composite
stellar populations of different mean temperatures, the use of 
different indices among those listed in Table 5 will be more
appropriate in each case. For a very metal-rich and cooler stellar
population of a massive elliptical galaxy, EW$_{8400}$ would be more
sensitive to changes in line strength, whereas for a metal-poor
globular cluster the bluer indices are more suitable.

Fig. 8 shows the $\Delta$v=+1 vibrational sequence of the  $\delta$
system, fitted to HR 3816. The features are clearly visible, although
much weaker than the bluer ones, even for such a low temperature as
T$_{\rm eff}$ = 2750 K. The behavior of the corresponding
index EW$_{9800}$ vs. T$_{\rm eff}$ (Fig. 9) indicates that this band
essentially does not  saturate in the temperature range considered.

\subsection{Integrated spectrum of NGC 6553}

We concentrate our attention on the bulge cluster NGC 6553, for which we
have a wealth of information: {\it i)} V and I photometry from the
Hubble Space Telescope (Ortolani et al. 1995); {\it ii)} J and K
photometry from the ESO 2.2m telescope (Guarnieri et al. 1998), resulting in
reliable determinations of reddening and distance to the cluster; {\it
iii)} detailed abundance analyses of stellar members (Barbuy et al.
1998), giving [Fe/H]$\sim$--0.4 and [$\alpha$/Fe]$\sim$+0.4;
{\it iv)} medium-resolution spectra of stars from the tip of the red
giant branch (Sec. 2) in the NIR, including several TiO bands, and {\it v)}
the integrated cluster spectrum covering the  spectral
interval $\lambda\lambda$ 6000 -- 9700 {\rm \AA}.

From the observed CMD, it is possible to compute the integrated
synthetic spectrum of NGC 6553 in a straightforward manner.  Such
method was employed by de Souza et al. (1993), Barbuy (1994) and Milone
et al. (1995) where synthetic spectra were used, and Santos et
al. (1995) where the stellar library of Jacoby et al. (1984) was
adopted.  On the other hand, the integrated spectrum of NGC 6553 was
reproduced in Bruzual et al. (1997), by using isochrones and several
spectral libraries (synthetic and observed).

The CMD of NGC 6553 was transformed into a $\log g\times$T$_{\rm eff}$
diagram, using the T$_{\rm eff}\times\log g\times$[Fe/H]$\times$(V--I)
relations from BCP98 for all the stars, except the ones
at the RGB tip. For the latter,  we adopted the T$_{\rm
eff}$ scale derived in Section 4. For the stars below the turn-off, we
adopted a Salpeter IMF plus the M-dwarf theoretical main sequence from
Baraffe et al. (1995).  Each star in the $\log g\times$T$_{\rm eff}$
plane has its absolute I magnitude (computed from distance and reddening
estimated by Guarnieri et al. 1998), and a corresponding synthetic
spectrum, which is chosen to be the one with closest atmospheric
parameters.  The integrated synthetic spectrum is then obtained by
summing up all the selected individual stellar spectra, weighted by
M$_I$. In Fig. 10 the result is compared with the observed integrated
spectrum of NGC 6553, adopting [M/H] = 0.0.

From the above calculations, we can estimate the contribution of stars
from all evolutionary stages to the integrated light of the cluster.
This is shown in Fig. 11, where it can be seen that the few stars
in the RGB tip largely dominate the cluster light in the
spectral region under consideration.

We note that the implementation of this grid of synthetic spectra to the code
of evolutionary populations synthesis by Bruzual \& Charlot (1998) is under way.

\section{ CONCLUSIONS}

We present a new grid of synthetic spectra in the interval
$\lambda\lambda$ 6000--10200{\rm \AA}. This grid is computed by
employing state of the art model atmospheres and has a higher
resolution than the ones currently in use in stellar population
synthesis in the same spectral interval.

A fit to NIR spectra of stars from the
RGB tip of NGC 6553 enabled us to derive a T$_{\rm eff}$ scale
for the M giants, based on the  TiO 
$\gamma$ system band intensities, on the one hand, and also to determine
 a set of semi-empirical electronic oscillator strengths for the
$\gamma$',
$\delta$, $\epsilon$ and $\phi$ systems of the TiO molecule. 

The adequacy of our spectral library for stellar population synthesis
purposes was tested for the case of NGC 6553, a well studied bulge 
globular cluster, for
which detailed abundance analyses of individual stars and well-defined
CMDs are available. In this framework, we obtained a good match to the
ISED of this cluster. This result warrants the use of our spectral grid
in stellar population synthesis studies in the NIR.

\acknowledgements                     
The authors are indebted to Robert Kurucz, France Allard and Bertrand
Plez for providing their model atmospheres, Roger Bell for the
theoretical TiO line list and for helpful comments, and to Michael
Bessell for making available his results in advance of publication. 
The anonymous referee is acknowledged for his helpful suggestions. All
the computations were carried out in a DEC-Alpha 3000/700 workstation.
RPS acknowledges the FAPESP PhD fellowship n$^o$ 93/2177-0.

\begin{deluxetable}{lccccccc}
\tablenum{1}
\tablewidth{0pt}
\tablecaption{ Log of the observations of field and cluster M giants. }
\tablehead{
\colhead {star} &
\colhead {V} &
\colhead{ $\alpha_{1950}$} & 
\colhead{$\delta_{1950}$} & 
\colhead{date} &
\colhead{exp.(m)} & 
\colhead{$\lambda\lambda$ ({\rm\AA}})  }
\startdata
 HR 625 &6.10 & 02$^{\rm h}$06$^{\rm m}$23.4$^{\rm s}$ & -18$^{\rm o}$00'55.5"
 &05.08.96 &2 & 5900-9200  \nl
 HR 3816 & 6.10   &09 30 59.2  & -62 34 01 &  01.02.96 &0.5 & 8090-10040  \nl
 NGC 6553-int. & -- & 18 05 11  & -25 55 06 & 19.07.96 & 30 
& 5900-9200   \nl
 NGC 6553-291 & 16.92 & "  & " & 10.06.97 & 30 
& 6865-8251   \nl
 NGC 6553-125& 15.89    & " & " & " & " & 6865-8905  \nl
 NGC 6553-288& 15.91   & " & " & " & " & "  \nl
 NGC 6553-47&  15.35    & " & " & " & " & "  \nl
 NGC 6553-118& 15.70    & " & " & " & " & "  \nl
 NGC 6553-211& 15.61     & " & " & " & " & "  \nl
\enddata
\end{deluxetable}

\begin{deluxetable}{lcccccccccccccccc}
\tablenum{2}
\tablewidth{0pt}
\tablecaption{Frame coordinates, observed (V-K) colours and temperatures of 6 
cool giants of NGC 6553.
Coordinates correspond to the frame in K band
 shown in Fig. 1 by Guarnieri et al.
(1998); star numbers and colours are also from Guarnieri et al.
The field stars HR 625 and HR 3816 are also included in the Table.
Relations colour-temperature by Bessell et al. (1998) - BCP
(theoretical/empirical), 
Lejeune et al. (1998) - LCB98, and Ridgway et al. (1980)
 are used. Gravities adopted are given in the last column. }
\tablehead{
\colhead {star} &
\colhead { X} &
\colhead { Y} &
\colhead {(V-K)} &
\colhead { T$_{BCP}$} &  
\colhead {T$_{LCB}$} &
\colhead {T$_{Ridgway}$} &
\colhead {T$_{TiO}$} &
\colhead { log g} }
\startdata
291/V12  & -4.7&324.0&9.75&  3250/3075& 3100 & 3060   &  3200& 0.0 \nl
125   &30.1    &276.5&7.21&  3520/3476 &    3550 & 3530&  3500& 0.7 \nl
288   &10.8    &334.0&7.93&  3420/3367 &    3400 & 3410&  3400& 0.5 \nl
 47   &63.7    &286.2&6.96&  3550/3518 &    3590 & 3570&  3400& 0.5 \nl
118   &30.0    &281.0&7.15&  3530/3476 &    3550 & 3540&  3500& 0.7 \nl
211   &14.1    &224.9&3.89$^*$&   --&           -- &  --   & 3500& 0.7 \nl
HR 625   &--    &--& 4.48 &   3660/3670 & 3700 & 3680 & 3600& 1.0 \nl
HR 3816 &-- & --& -- &   --& -- &  --   & 2750& -0.5 \nl
\enddata
\tablecomments{$^*$ For the star 211 the (V-K) colour appears to be
in error, since the spectrum is incompatible with a high temperature}
\end{deluxetable}

\begin{deluxetable}{lcccccccccc}
\tablenum{3}
\tablewidth{0pt}
\tablecaption{Electronic oscillator strengths from the literature
(Davis, Littleton \& Phillips 1986 - DLP86;
Brett 1990 - B90; Jorgensen 1994 - J94;
Hedgecock, Naulin \& Costes 1995 - HNC95;
Langhoff 1997 - L97; Alvarez \& Plez 1998 - AP98)
compared to the value found in the present work}
\tablehead{
\colhead {system} &
\colhead { DLP86} &
\colhead { B90} &
\colhead { J94} &
\colhead { HNC95} &
\colhead { L97} &
\colhead { AP98} &
\colhead { This Work} }
\startdata
$\gamma$' & 0.138  &  ---  &  0.14 & 0.093 & 0.108 & 0.0935 & 0.06 \nl
$\gamma$  & 0.165  & 0.22  &  0.15 & 0.079 & 0.092 & 0.0786 & 0.12   \nl
$\delta$  & 0.05   & 0.05  &  0.048& ---   & 0.096 & 0.048 & 0.05   \nl
$\epsilon$ & ---   & 0.006 &  0.014& $<$0.0056 &  0.002 & 0.0023 & --   \nl
$\phi$    & 0.05  & 0.05   &  0.052 & ---  & 0.018 & 0.0178 & 0.01   \nl
\enddata
\end{deluxetable}

\begin{deluxetable}{lccccccc}
\tablenum{4}
\tablewidth{0pt}
\tablecaption{Stellar parameters 
and corresponding photospheric models used to build
the grid of synthetic spectra; models are from
Kurucz (1992) - K92, Plez (1997) - P97, Plez et al.
(1992) - PBN92 and Allard \& Hauschildt (1995) - AH95}
\tablehead{ 
\colhead{T$_{\rm eff}$/step} &
\colhead{log g/step} & 
\colhead{[Fe/H]}  & 
\colhead{models}  }
\startdata
 4500-6000/250 & 0.0-5.0/0.5 & -0.3,0.0 & K92   \nl
4000-4500/250 & 2.0-5.0/0.5 & -0.3,0.0 & K92   \nl
4000-4500/250 & 0.0-1.5/0.5 & -0.6,-0.3,0.0 & P97 \nl
3600-4000/200 & -0.5-1.0/0.5  & -0.6,-0.3,0.0  & P97  \nl
3000-3400/200 & -0.5-1.5/0.5 & -0.6,-0.3,0.0 & PBN92  \nl
2750-3000/250 & -0.5-1.5/0.5 & -0.6,-0.3,0.0 & " \nl
2500    & -0.5 & -0.6,-0.3,0.0 & " \nl
2800-3200/200 & 3.5-5.0/0.5 & -0.5,0.0 & AH95 \nl
3300-3700/200 & 3.5-5.0/0.5 & -0.5,0.0 & " \nl
2500 & 5.5 & -0.5,0.0 & " \nl
\enddata
\end{deluxetable}

\begin{deluxetable}{lccccccccc}
\tablenum{5}
\tablewidth{0pt}
\tablecaption{Definition of spectral indices measured}
\tablehead{
\colhead {Index} &
\colhead {T$_{eff}$ range} &
\colhead { Blue continuum} &
\colhead { Bandpass} &
\colhead { Red continuum} }
\startdata
EW$_{6000}$ & $\leq$ 3600& 6143.95-6146.41 & 6145.18-6540.93 & 
6539.58-6542.27\nl
EW$_{6000}$ & $\geq$ 3600& 6067.47-6074.81 & idem & idem \nl
EW$_{6600}$ & $\leq$ 3600& 6539.58-6542.27 & 6540.93-7048.22 & 
7548.70-7553.14\nl
EW$_{6600}$ & $\geq$ 3600& idem & idem & 7594.40-7598.00\nl
EW$_{7100}$ & $\leq$ 3600& 7046.36-7050.08 & 7048.22-7550.92 & 
7548.70-7553.14\nl
EW$_{7100}$ & $\geq$ 3600& 6539.58-6542.27 & idem & 7594.40-7598.00\nl
EW$_{7600}$ & $\leq$ 3600& 7548.70-7553.14 & 7550.92-8190.40 & 
8188.86-8191.93\nl
EW$_{7600}$ & $\geq$ 3600& 7593.64-7598.33 & idem & 8259.46-8261.50\nl
EW$_{8400}$ & $\leq$ 3600& 8188.86-8191.93 & 8190.40-8853.40 & 
8850.67-8856.13\nl
EW$_{8400}$ & $\geq$ 3600& 8259.46-8261.50 & idem & idem \nl
EW$_{9800}$ & ---  & 9721.67-9725.00 & 9726.00-9972.00&
   9975.23-9981.52\nl
TiO$_2$(Burstein) & ---& 6069.00-6142.75 & 6192.00-6273.25 & 
6375.00-6416.25\nl
TiO$_2$(Worthey) & ---& 6068.375-6143.375 & 6191.375-6273.375 & 
6374.375-6416.875\nl
\enddata
\end{deluxetable}

\clearpage

\centerline {\bf FIGURE CAPTIONS}

\figcaption{Strongest TiO bands in the near infrared, corresponding
to the $\gamma$, $\gamma$', $\epsilon$ and $\delta$ electronic systems.
 \label{fig1}}

\figcaption{$\gamma$ and $\gamma$' bandheads for HR 625. Solid lines:
synthetic spectrum computed for
T$_{\rm eff}$=3600 K, log g = 1.0, [Fe/H] = 0.0;
dashed line: observed spectrum
 \label{fig2}}

\figcaption{$\Delta$v=0 and $\Delta$v=-1 vibrational sequences
 of the $\gamma$ system
for the star NGC 6553:288. Solid line: synthetic spectrum computed
with T$_{\rm eff}$=3400 K, log g = 0.5, [Fe/H] = 0.0;
dashed line: observed spectrum.
 \label{fig3}}

\figcaption{$\Delta$v=2 vibrational sequence of the $\gamma$ system,
$\Delta$v=0 of the $\epsilon$ system and $\Delta$v=0 of the
$\delta$ system for NGC 6553:288. Solid line: synthetic spectrum
computed with T$_{\rm eff}$=3400 K, log g = 0.5, [Fe/H] = 0.0;
dashed line: observed spectrum.
 \label{fig4}}

\figcaption{TiO features vs. T$_{\rm eff}$ for [Fe/H] = -0.3 and
log g = -0.5, 0.0, 0.5, 1.0 and 1.5.
 \label{fig5}}

\figcaption{TiO features vs. T$_{\rm eff}$ for log g = 1.5 and
[Fe/H] = -0.6, -0.3 and 0.0.
\label{fig6}}

\figcaption{Synthetic spectra for 
Teff = 3400 K, log g = 0.5 and [Fe/H] = 0.0,
where the TiO$_2$ index as defined  
according to Worthey et al. (1994), and our EW$_{6000}$ index
are indicated.
\label{fig7}}

\figcaption{$\Delta$v = +1 vibrational sequence of the $\delta$ system
in HR 3816. Solid line: synthetic spectrum computed for 
T$_{\rm eff}$=2750 K, log g = -0.5, [Fe/H] = 0.0; dashed line: observed
spectrum.
\label{fig8}}

\figcaption{TiO feature EW$_{9800}$ vs. T$_{\rm eff}$. Symbols are
as in Fig. 5 and open stars correspond to dwarfs.
\label{fig9}}
 
\figcaption{Integrated spectrum of NGC 6553. Dashed line:
observed spectrum; solid line: synthetic spectrum computed 
as described in Sec. 5.1.
\label{fig10}}

\figcaption{Contribution of stars in different evolutionary stages
to the integrated light of NGC 6553.
\label{fig11}}

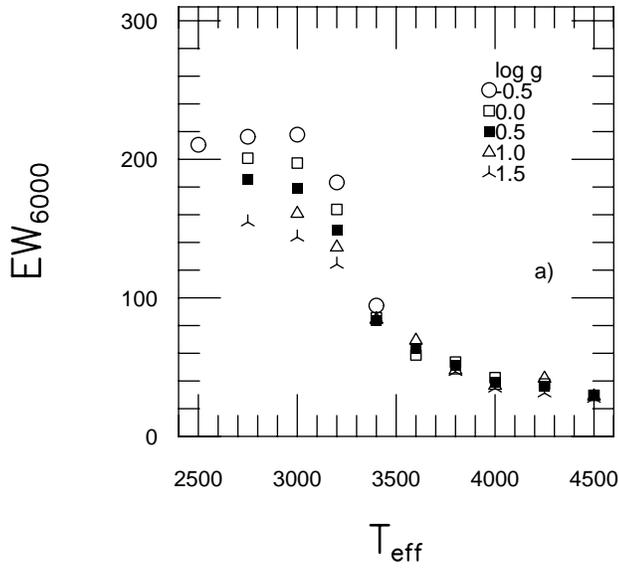
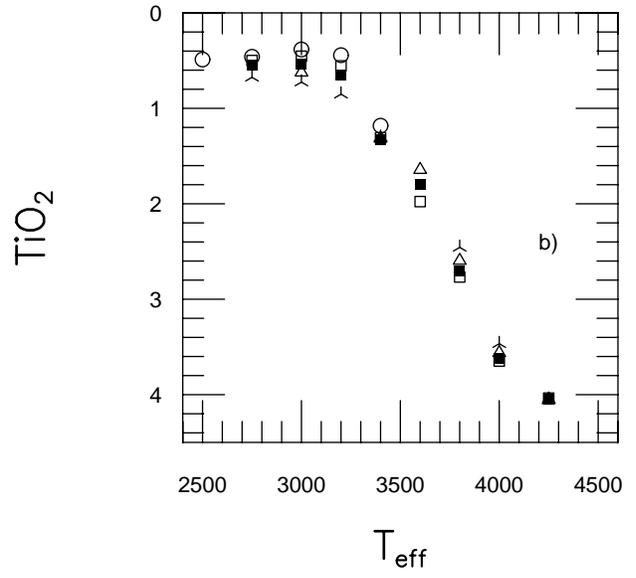
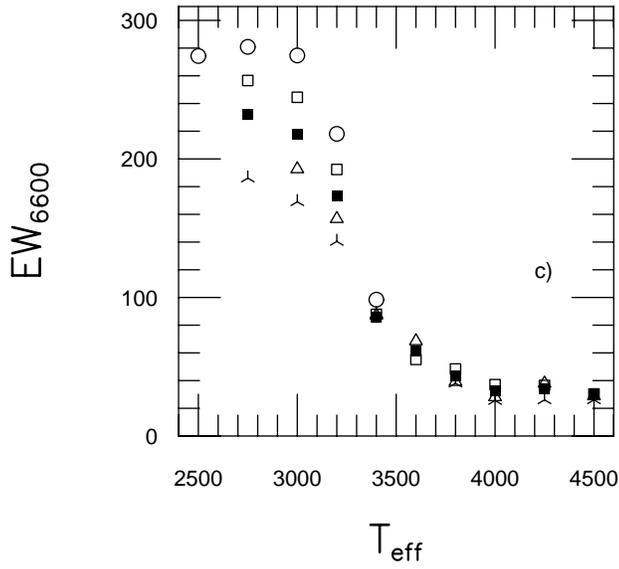
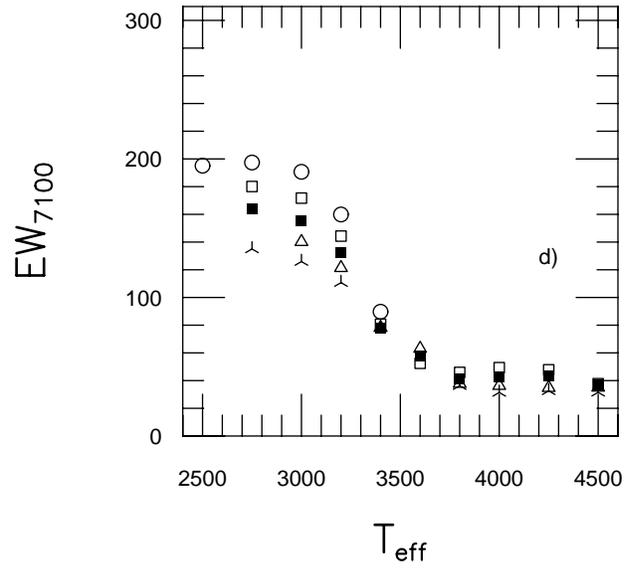
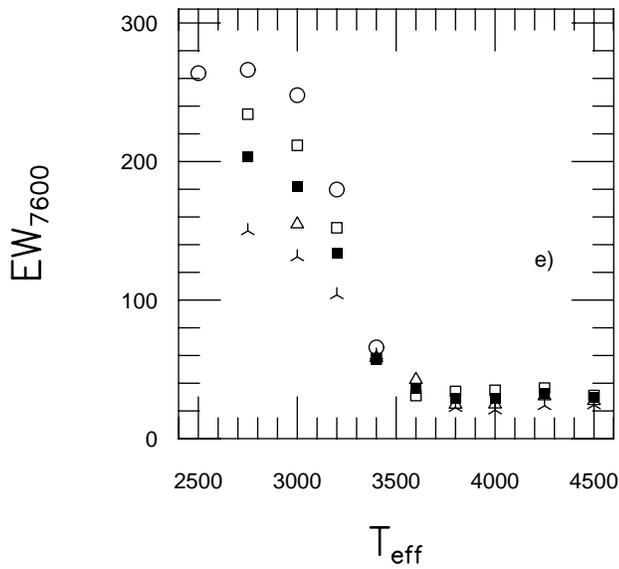
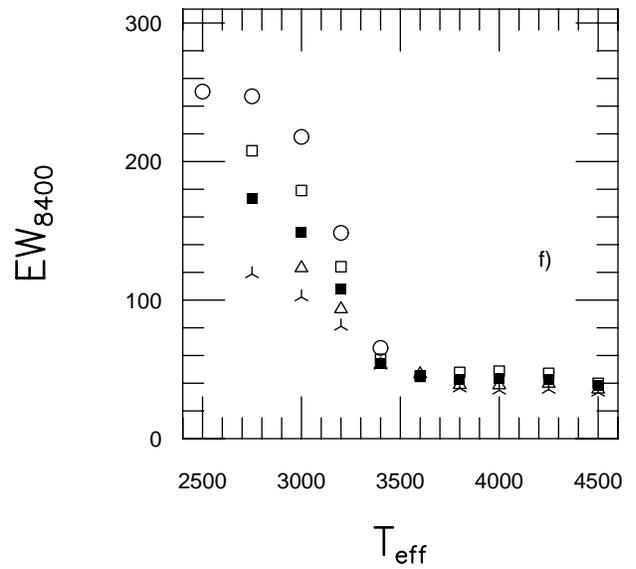
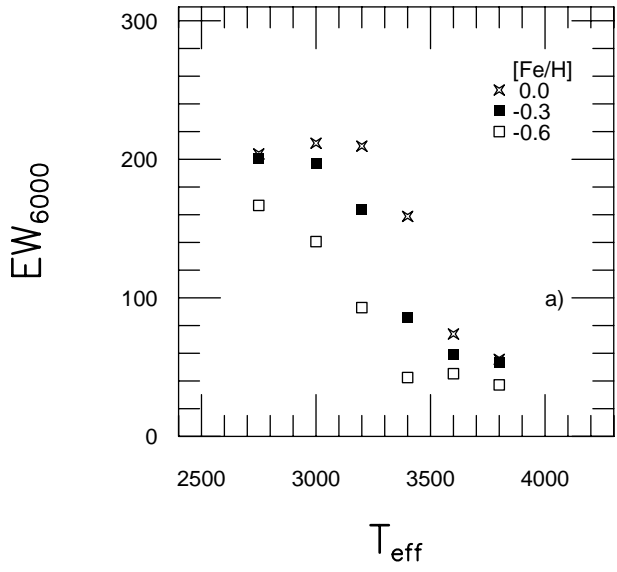
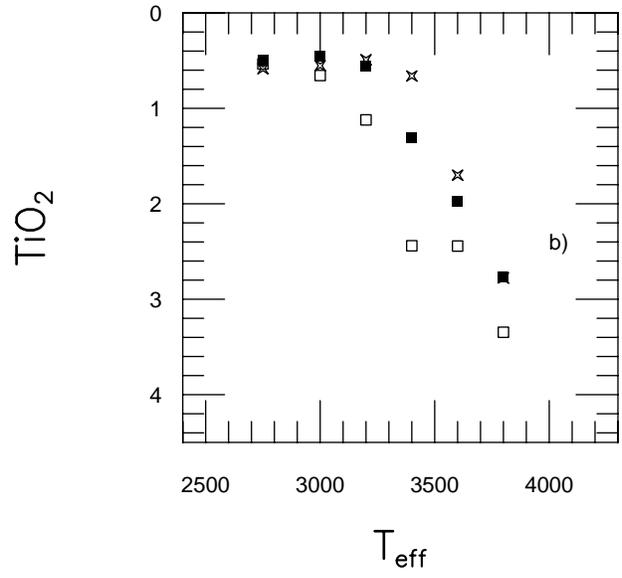
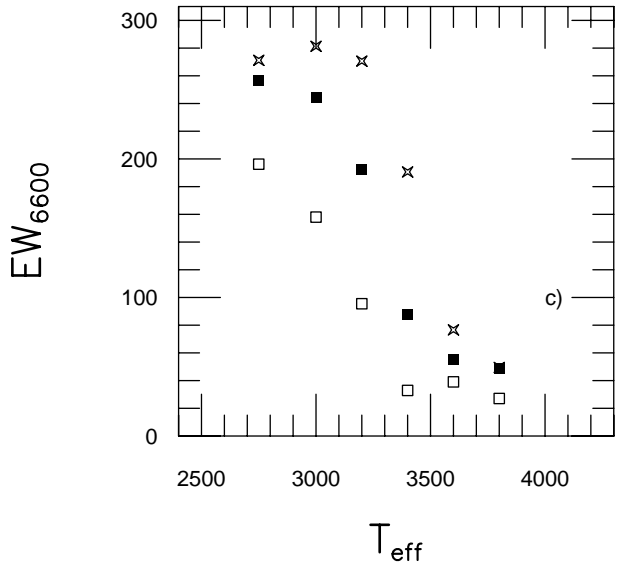
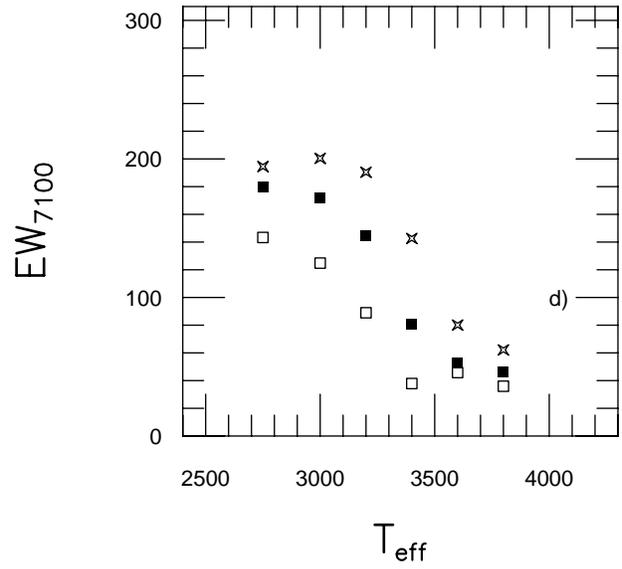
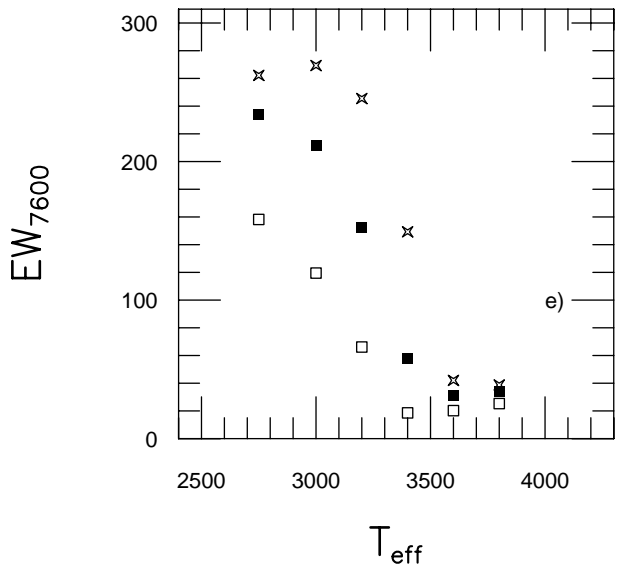
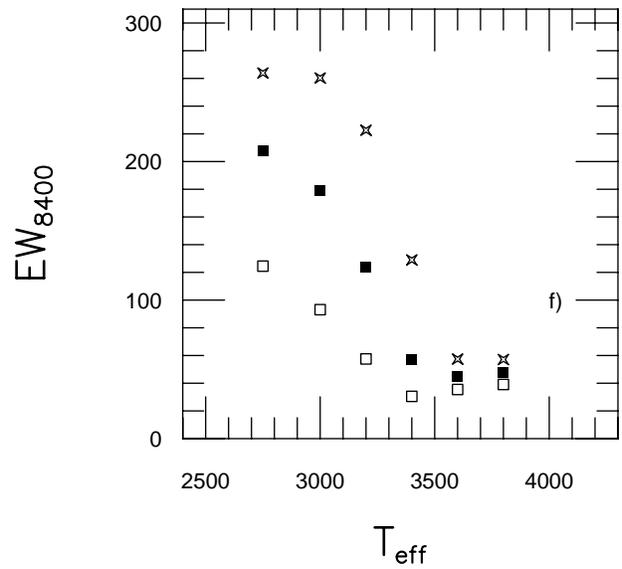
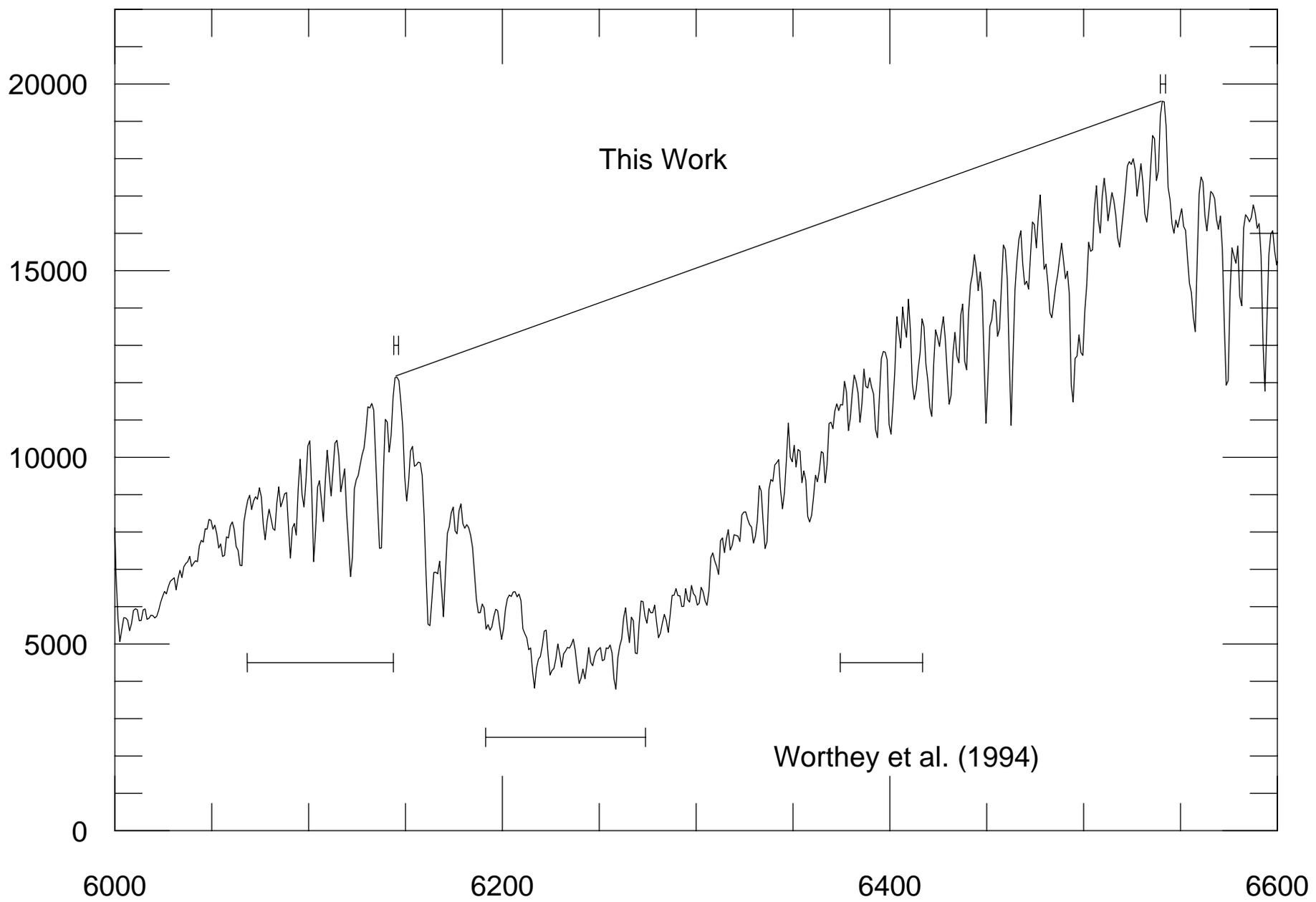
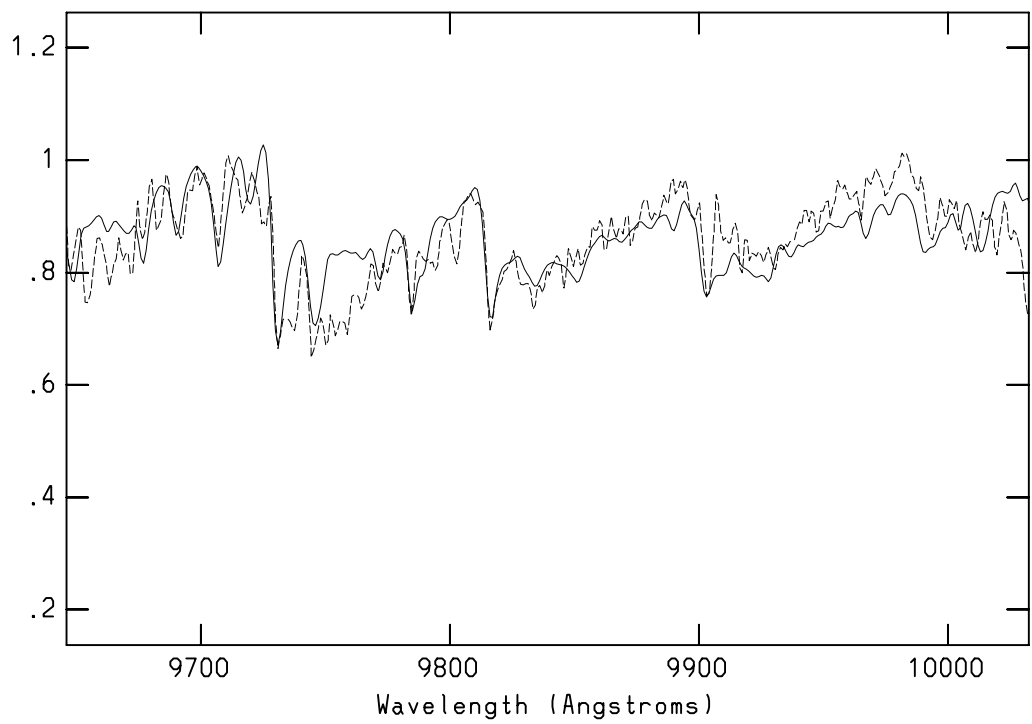
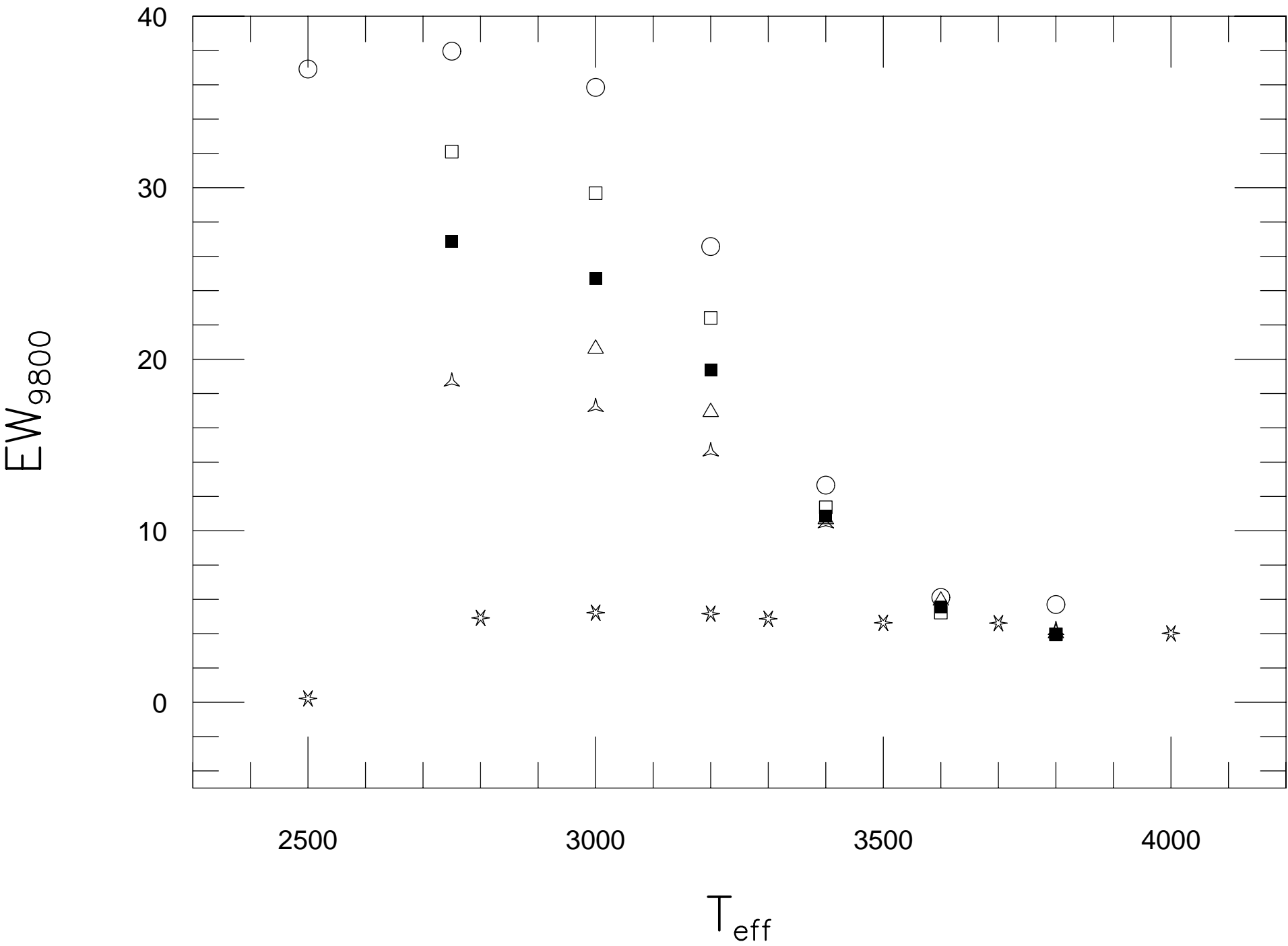
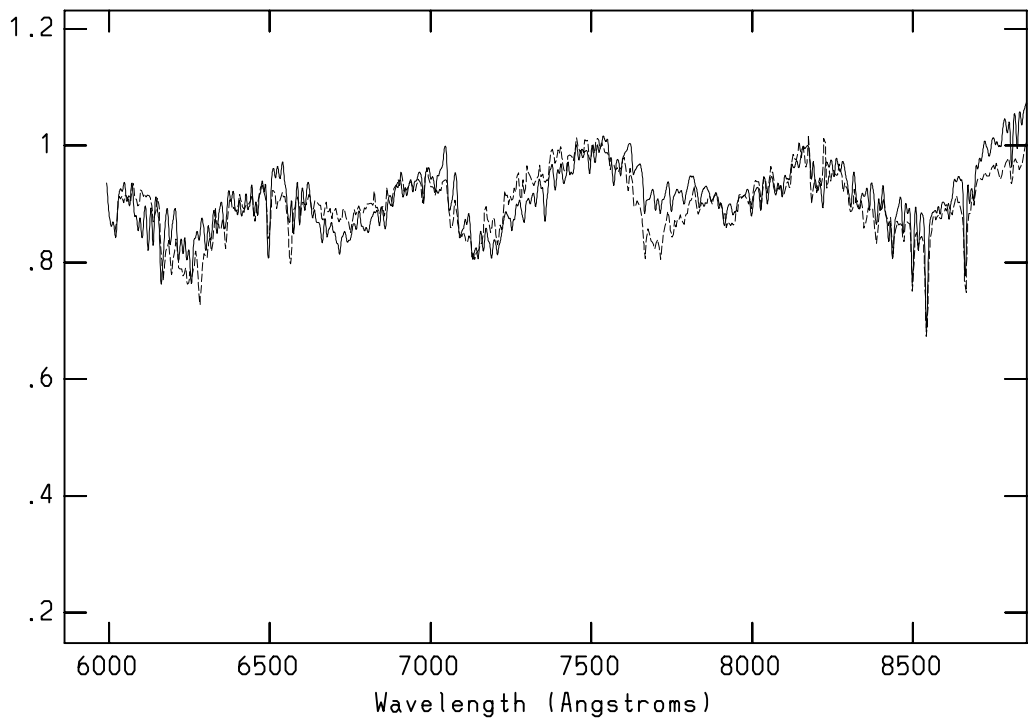
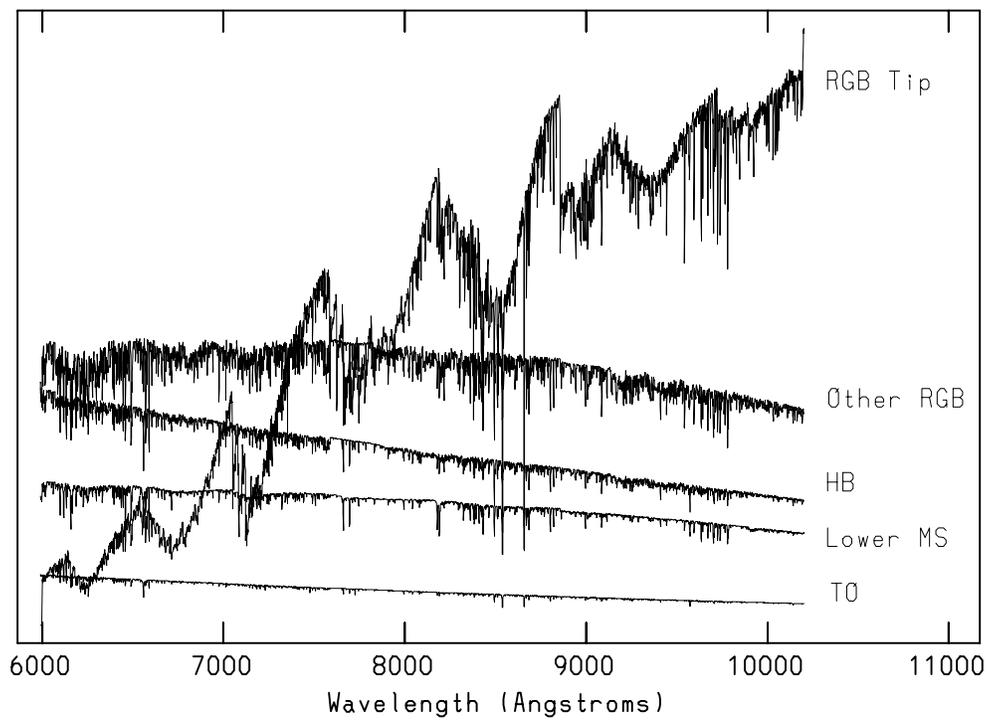
\end{document}